\documentclass[sigconf]{acmart}
\usepackage{indentfirst}
\usepackage{url} % Include the URL package
\usepackage{hyperref} % Optional, for clickable links
\usepackage{float}
\acmConference[]{}{}{}
\acmYear{2024}
\acmPrice{}
\acmBooktitle{}
\acmISBN{}
\usepackage[utf8]{inputenc}
\usepackage{geometry}
\usepackage{booktabs} % For better table formatting
\usepackage{amsmath}  % For mathematical symbols
\usepackage{siunitx}  % For proper number formatting
\usepackage{caption}
\acmDOI{}
\AtBeginDocument{%
  \providecommand\BibTeX{{%
    \normalfont B\kern-0.5em{\scshape i\kern-0.25em b}\kern-0.8em\TeX}}}

\begin{document}

\title{SPACEX: Exploring metrics with the SPACE model for developer productivity}

\author{Sanchit Kaul}
\email{skkaul@ucdavis.edu}
\affiliation{
  \institution{University Of California, Davis}
  \country{}
}

\author{Kevin Nhu}
\email{knhu@ucdavis.edu}
\affiliation{
  \institution{University Of California, Davis}
  \country{}
}

\author{Jason Eissayou}
\email{jdeissayou@ucdavis.edu}
\affiliation{
  \institution{University Of California, Davis}
  \country{}
}

\author{Ivan Eser}
\email{eeser@ucdavis.edu}
\affiliation{
  \institution{University Of California, Davis}
  \country{}
}

\author{Victor Borup}
\email{vborup@ucdavis.edu}
\affiliation{
  \institution{University Of California, Davis}
  \country{}
}

\maketitle

\section{Introduction \& Motivation}
%\textit{Explain the purpose of your study, why the topic is important, and the key problem or question you aim to address. }

Measuring a developer's productivity may initially seem like an easy and simple task. However, this is a very crucial topic as productivity is a driving force in the software industry and often helps companies maintain a competitive edge in the market. Traditionally, the metrics used by company managers to measure productivity may include total commit count and code churns (sum of added and removed lines of code). However, research has shown that the definition of developer productivity actually carries much more complexity, leading to the development of multiple frameworks to measure productivity. One of these frameworks is the Blumberg Model of Performance, developed in 1982, which attributes a developer's performance to three dimensions (Capacity, Willingness, and Opportunity) \cite{BlumbergPringle}.

A more recent framework and the main reference of our research project is the SPACE Framework, which consists of five dimensions (Satisfaction, Performance, Activity, Communication, and Efficiency) introduced by Microsoft in 2021. Unlike traditional metrics, SPACE considers both qualitative and quantitative aspects in order to measure a developer's productivity. The SPACE Framework developers believe that individual metrics do not completely measure developer productivity, but rather a combination of multiple dimensions within the framework can capture productivity in a more meaningful manner \cite{acmpaper}.

Despite extensive studies and research done, there is little consensus for a proper metric to measure developer productivity as its definition may differ between individuals and teams \cite{Storey2022ProductivityQuality}. Many project managers still rely on simplistic and traditional metrics such as lines of code and commit counts \cite{lima2015assessing} to justify how to reward or train their employees. This method is not as reliable since it does not capture qualitative properties such as communication, satisfaction, or efficiency. To address this issue, our study utilizes the SPACE Framework to determine how many of these five dimensions can be measured using repository mining and develop a holistic metric by combining multiple framework dimensions using both qualitative and quantitative metrics to provide a more effective (and accurate) measure of productivity.

\section{Background}
\subsection{Literature Review}

% \textit{Summarize existing studies, theories, or findings that are relevant to your research. Highlight gaps or limitations in the literature that your study addresses.}

Measuring developer productivity has been the topic of many related studies in empirical software engineering. Typically, traditional metrics, such as commit counts and lines of codes (LOC) have been used, but these generally do not capture important (and qualitative) aspects of the development process such as code quality, collaboration, and well-being of developers. To account for these shortcomings, Microsoft researchers created the SPACE Framework in 2021 to evaluate productivity across five different dimensions, including Satisfaction, Performance, Activity, Communication, and Efficiency. Unlike traditional approaches, SPACE directly acknowledges how it is not possible for a singular metric to fully encapsulate developer productivity. Our study strives to build upon this framework by creating a holistic metric that incorporates multiple factors from multiple dimensions using repository mining.

In 2015, Lima et al. (2015) explored the use of repository mining-based metrics, such as LOC (lines of code), method complexity, and bug resolution. Their findings revealed a major issue in terms of over-reliance on a singular metric. Doing so will often lead to inaccurate conclusions since developer contributions can differ greatly based on experience, project structure, and assigned tasks \cite{lima2015assessing}. Additionally, a large-scale GitHub study on programming languages and software quality revealed that different programming languages may have some effect on the quality of software, but their actual impact is minimal compared to other factors \cite{RayPaper}. Both of these studies emphasize the idea that productivity cannot be accurately depicted through a single metric, which is why we intend for our research to take a multi-dimensional approach by integrating multiple SPACE dimensions to provide a more balanced measure of developer productivity.

Microsoft teams were also examined and studied for how they reuse code to improve efficiency, identifying five common strategies in the process, including personal analysis reuse, personal utility libraries, team-shared analysis code, team-shared template notebooks, and team-shared libraries. While this research did shed light on potential workflow improvements, it lacked an explanation in quantifying the actual impact these strategies had on productivity or comparing them to more traditional metrics \cite{EppersonPaper}. Additionally, the research appears to be rather domain-specific, focusing specifically on Microsoft employees, thus leading to a generalizability concern. Our work aims to bridge this gap by using repository mining to analyze productivity trends across a broader range of teams and projects. 

The Delta Maintainability Model quantifies the associated risk with code changes using complexity and coupling metrics. While this is useful in determining maintainability, it unfortunately disregards other aspects of developer performance highlighted by the SPACE Framework, such as efficiency and collaboration \cite{biase}. DMM is a valuable metric that can be used to measure developer performance, as higher DMM scores are indicative of worse performance scores (thus signifying an inverse relationship between DMM and performance). An implementation of DMM is available within the PyDriller library, and includes Unit Size, Unit Complexity, and Unit Interfacing metrics, providing values on a scale between 0.0 to 1.0, with values closer to 0.0 indicating low risk and values closer to 1.0 indicating high risk. 

In a more recent study, LLMs were used in a sentiment analysis study directly comparing bigger large language models (bLLMs) and smaller fine-tuned LLMs (sLLMs). Using five software engineering datasets, it was determined that bLLMs exhibited superior performance in zero-shot and few-shot learning scenarios, where there is limited labeled data. sLLMs, on the other hand, performed better when ample training data was available \cite{ZhangPaper}. However, the main limitation of this study is, once again, its lack of generalizability, choosing to focus specifically on sentiment analysis rather than developer productivity as a whole. While useful in determining satisfaction, sentiment analysis does not account for other broader productivity factors such as efficiency, performance, or teamwork. To address this limitation, our research integrates sentiment analysis with repository mining, allowing us to analyze satisfaction in relation to other indicators of productivity.

\subsection{Research Questions}
%\textit{Clearly state the specific research questions or hypotheses your study seeks to explore or test.}

\begin{enumerate}
\item How can a holistic metric be defined within the SPACE framework to effectively measure developer productivity?
\item How does this holistic metric compare to traditional metrics such as the number of commits or lines of code added?
\item In what specific scenario or contexts does the holistic SPACE metric capture a property of developer productivity that traditional metrics might miss? (For example, code quality may not be represented by total commits, an Activity metric.)
\end{enumerate}

\section{Methodology}

\subsection{Data}

For the data collection, we used PyDriller and GitHub's REST API to mine information from around 100 repositories, meticulously selecting projects with around 3-10 active contributors and consistent activity over the past 6 months. The dataset was intended to be used for a variety of SPACE dimensions. Primary data cleaning included (but is not limited to) de-aliasing (removing duplicate authors or contributors), removing bot accounts, and inactive authors. Within the larger data set, different metrics are used to measure the different dimensions. While this dataset is large, it is comprised solely of open-source Github projects, which may bias the data as compared to projects in a more structured or corporate setting.

\subsection{Satisfaction}

\subsubsection{Background}

In past research, developer satisfaction was always measured using direct surveys with the developers.
An example of this can be found in research done by Smit et al., where they did a Likert scale survey
and posed this question to the developers: \emph{``As a developer, my job satisfaction has increased or will increase due to the use of GitHub Copilot.''}
Developers could select between ``Strongly agree,'' ``Agree,'' ``Neither agree nor disagree,''
``Disagree,'' and ``Strongly disagree.'' \cite{smit2024copilot}

For the scope of our research, we were unable to send out surveys to the thousands of developers that we
mined information from. Due to this limitation, we turned to a different method of measuring developer satisfaction,
which consisted of looking at the commit and issues messages. This was the closest we could get personally to the developer,
as these commit and issue messages can really show how a developer is feeling at that point in time.

After data collection, sentiment analysis was performed on each individual commit message. Each message was classified
as having positive, negative, or neutral sentiment using a RoBERTa-based model, pre-trained for sentiment classification \cite{CardiffNLP}.
Following classification, we looked mainly at the messages with negative sentiments. For each author/project combination,
the percentage of negative commit messages was calculated by dividing the number of negative messages by the total number
of commits for that author/project, then multiplying by 100. This calculated percentage was then added as a new column
alongside its associated unique combination of author and project name.

\subsubsection{Variables/Metrics to Consider}

For measuring Satisfaction, these variables were used:
\begin{itemize}
    \item \texttt{Total.Commits}: The total number of commits from each author to the project's codebase.
    \item \texttt{Project.Age..Years.}: How long a project was worked on, from the first commit to the last commit.
    \item \texttt{Negative.Commit.Percentage}: The percentage of commits classified as ``negative'' by a
          trained RoBERTa transformer model.
    \item \texttt{Total.Issues}: The total number of issues reported.
    \item \texttt{Total.PRs}: The total number of pull requests submitted.
    \item \texttt{Project.Name}: The name of the project.
\end{itemize}

All authors with fewer than 20 commits were excluded to mitigate the influence of low-activity one-off committers.

\subsubsection{Methods for Statistical Analysis}
Initially, for the statistical analysis, we looked at using a linear mixed model (LMM) with the \texttt{lmer} function
to analyze the relationship between negative commit percentage and total commits. While we log-transformed
the dependent variable (\texttt{Total.Commits}) to improve normality, checking the residuals of the fitted LMM showed
major violations of the normality assumption. The quantile-quantile (Q-Q) plot of the residuals showed an S-shaped pattern,
indicating that the residuals were not normal. This deviation from normality is not good when using LMMs because the
validity of p-values and confidence intervals relies on normally distributed residuals.

One reason we believe this happened is that \texttt{Total.Commits}, representing count data, still exhibited skewness.
Even after removing authors with fewer than 20 commits, the mean count remained skewed toward the lower end.
A log transformation of \texttt{Total.Commits} did not fully resolve the non-normality, as confirmed by the S-shaped Q-Q plot.

Because of these two issues (violation of the normality assumption and the dependent variable being skewed),
we decided to use a Generalized Linear Mixed Model (GLMM). The reason for this is that GLMMs are specifically designed to handle non-normal
response variables which includes count data.

One main problem we encountered with the GLMM was the warning:
\emph{``Model is nearly unidentifiable: very large eigenvalue.''}
This indicated some numerical instability due to our predictor variables having widely varying scales.
To address this, we standardized all continuous predictors by using R’s built-in \texttt{scale()} function, transforming
each variable so that it has mean 0 and standard deviation 1 (z-scores).

Once all variables were correctly scaled, our main GLMM model included:
\begin{itemize}
    \item \texttt{Negative.Commit.Percentage\_scaled} (primary predictor of interest)
    \item \texttt{Project.Age..Years.\_scaled} (control variable)
    \item \texttt{Total.Issues\_scaled} (control variable)
    \item \texttt{Total.PRs\_scaled} (control variable)
\end{itemize}
as fixed effects, and a random intercept for project name, \texttt{(1 | Project.Name)}. In R, the model was specified as:

\begin{verbatim}
model_poisson_scaled <- glmer(
    Total.Commits ~ Project.Age..Years._scaled +
                     Negative.Commit.Percentage_scaled +
                     Total.Issues_scaled +
                     Total.PRs_scaled +
                     (1 | Project.Name),
    data = finalRemovedLowCommits_scaled,
    family = poisson
)
\end{verbatim}

In terms of model diagnostics, we assessed multicollinearity among the fixed effects using variance inflation factors (VIFs) from the
\texttt{car} package (\texttt{vif()}). All VIFs were below 1.7, indicating no serious multicollinearity.
We also examined a Q-Q plot of deviance residuals for deviations from normality. We observed a deviation
in the right tail (points curving upward), suggesting right-skewness. However, the Poisson GLMM is
better suited for count data and can accommodate such skew more effectively than a standard linear model.

\subsection{Performance}

\subsubsection{Background}

The original \textit{“SPACE of Developer Productivity”} paper defines the "Performance" dimension of the SPACE framework as the outcome of a system or process. In other words, one would ask if the code that a developer writes reliably accomplished what it was intended to do when evaluating performance. It can be rather difficult to quantify since it is difficult to associate individual contributions directly with product outcomes. For example, producing a large amount of code does not necessarily equate to producing high-quality code. As a result of this outcome-driven perspective of evaluating performance, the paper suggests focusing on things like software quality (i.e., fixing bugs) and impact (i.e., customer satisfaction) \cite{acmpaper}.

However, one major limitation is that we don't have much (if any) information to use and analyze in terms of the customer/consumer side of the software being developed, with the exception of GitHub issues. For the sake of simplicity, we focused primarily on evaluating software quality.

Mining and analyzing data for this particular dimension proved to be rather different from the other dimensions. Seeing as the original dataset contained information based on each commit for each of the 100 repos that were mined, this would not really prove very useful here since we are primarily interested in evaluating performance on a developer-by-developer basis for each repo. 

As a result, the data collection differed slightly here, as we shall see in the next section.

\subsubsection{Variables/Metrics to consider}

For the Performance dimension of the SPACE Framework, we were interested in evaluating Performance from both the repository level and the contributor level (i.e., focusing on each developer per repo).

Thus, the following metrics are considered both at the repository level and the contributor level:

\textbf{Repository-level Metrics}

\begin{itemize}
    \item \texttt{CI/CD Success Rate} (primary predictor of interest)
    \item \texttt{Average PR Merge Time} 
\end{itemize}

\textbf{Contributor-level Metrics}

\begin{itemize}
    \item \texttt{Total Commits} (primary predictor of interest)
    \item \texttt{Bug Fix Commits} (primary predictor of interest)
    \item \texttt{Code Churn} 
\end{itemize}

\subsubsection{Methods for Statistical Analysis}

The Exploratory Data Analysis (EDA) phase of the Performance study focused on understanding the distribution of key variables, identifying any lingering outliers, and finding preliminary relationships between factors, while the CDA phase involved formal hypothesis testing through multiple regression models and partial correlation analysis.

For the EDA phase, we first computed descriptive statistics for both repository-level and contributor-level datasets. Winsorization was applied to \texttt{avg\_pr\_merge\_time}, trimming extreme values at the 1st and 99th percentiles to reduce the influence of outliers. Additionally, log transformations were applied where necessary to address skewness in the data, specifically using a $\log(x + 1)$ transformation for \texttt{avg\_pr\_merge\_time}. While a logit transformation for \texttt{ci\_cd\_success\_rate} was considered, it was not applied by default due to the concentration of values near 1. Scatter plots with fitted regression lines were also created to visualize relationships between \texttt{ci\_cd\_success\_rate} and \texttt{avg\_pr\_merge\_time}, as well as \texttt{total\_commits} and \texttt{code\_churn}, and \texttt{bug\_fix\_commits} and \texttt{code\_churn}. To quantify these relationships, Pearson correlation coefficients were computed, providing an initial assessment of how strongly these variables were associated.

Building off of the EDA, the CDA phase uses multiple regression models and partial correlation analysis to formally test relationships between variables. At the repository level, a multiple regression model was fitted to predict \texttt{avg\_pr\_merge\_time} as a function of \texttt{ci\_cd\_success\_rate} and \texttt{total\_commits}. At the contributor level, two separate models were fitted: one predicting \texttt{code\_churn} based on \texttt{total\_commits}, and another predicting \texttt{code\_churn} based on \texttt{bug\_fix\_commits}. These models were assessed using standard regression diagnostics, including residual plots to check for linearity and normality. 

To further refine the analysis, partial correlation analysis was conducted to account for potential confounding effects. This was done by first regressing each predictor and response variable on the control variable (\texttt{code\_churn}), extracting the residuals, and then computing the Pearson correlation between the residuals. From here, we determined if relationships between variables persisted after controlling for the effects of code churn.

We developed the following hypotheses for both repository-level and contributor-level relationships:

\textbf{Repository-Level Hypotheses}

\begin{itemize}
    \item \textbf{H$_0$ (Null Hypothesis):} There is no significant relationship between CI/CD success rate and average PR merge time, controlling for total commits.
    \item \textbf{H$_1$ (Alternative Hypothesis):} CI/CD success rate has a significant effect on average PR merge time.
\end{itemize}

\textbf{Contributor-Level Hypotheses}

\begin{itemize}
    \item \textbf{H$_0$ (Null Hypothesis):} There is no significant relationship between total commits and code churn.
    \item \textbf{H$_1$ (Alternative Hypothesis):} Total commits are positively associated with code churn.

    \item \textbf{H$_0$ (Null Hypothesis):} There is no significant relationship between bug-fix commits and code churn.
    \item \textbf{H$_1$ (Alternative Hypothesis):} Bug-fix commits are positively associated with code churn. 
\end{itemize}

\subsection{Activity}

\subsubsection{Background}

The \textit{“SPACE of Developer Productivity”} paper defines the “Activity” dimension of the SPACE framework as the count of actions or outputs, making this specific dimension the most quantifiable dimension in the SPACE framework. As the paper states, no singular metric can be used to measure a developer’s productivity. Instead, multiple factors should be considered to make inferences. Traditionally, a developer’s activity is measured simply by the number of lines of code (LOC) written. However, researchers today have determined that this is not a very good indicator of activity, as the number of lines of code can vary depending on the programming language used—some are easier to write in one language than another \cite {acmpaper}.

For this reason, we aim to answer the following research question:

\begin{quote}
\textit{How does our model compare to traditional metrics such as the number of commits or lines of code added?}
\end{quote}

The goal of this specific part of the research is to assess how well the traditional metric of measuring activity through lines of code is predicted by an author’s total commits and other variables, such as average complexity per method, using a multivariate regression model.

\subsubsection{Variables/Metrics to Consider}

Since this study focuses on activity metrics, the primary features of interest include:
\begin{itemize}
    \item \texttt{Code Churn} – The sum of added and removed lines of code.
    \item \texttt{Total Commits} – The number of commits made by a developer.
    \item \texttt{Average Complexity per Method} – A measure of method complexity per developer.
\end{itemize}

\subsubsection{Methods for Statistical Analysis}

Following the data science pipeline, we first define the hypotheses for our analysis:

\begin{itemize}
    \item \textbf{Null Hypothesis ($H_0$)}: There is no relationship between code churn and total commits in terms of activity.
    \item \textbf{Alternative Hypothesis ($H_A$)}: There is some relationship between code churn and total commits in terms of activity.
\end{itemize}

The response variable for prediction is \texttt{Code Churn}, while predictor variables include \texttt{Total Commits} and \texttt{Average Complexity per Method}. To ensure meaningful inference, we normalized our dataset by applying a logarithmic transformation to the response variable. 

We employed a multivariate regression model to examine how well total commits and average complexity per method can predict code churn. Since multiple dependent variables are involved, a multivariate regression approach was deemed most appropriate.

\subsection{Communication/Collaboration}

\subsubsection{Background}

The “SPACE of Developer Productivity” framework introduced by Forsgren et al. (2021) \cite{acmpaper} defines the “Communication” dimension as the interactions and collaborations among developers, emphasizing its critical role in teamwork and enhancing productivity within software development teams. Communication is inherently challenging to measure directly through repository data, as much of it occurs in private channels such as Slack or Discord. However, the framework suggests that indirect metrics derived from repository mining can infer communication patterns, offering insights into collaborative behaviors that traditional metrics overlook.

This study addresses \textbf{research question \#3}: 
\begin{quote}
\textit{In what specific scenarios or contexts does the holistic SPACE metric provide unique insights into developer productivity that traditional metrics might miss?}
\end{quote}

Specifically, we aim to demonstrate how our communication metrics—\textit{Contributors Experience} and \textit{Commit Interaction Frequency (CIF)}—reveal collaboration dynamics that traditional metrics like commit counts fail to capture, such as the frequency and distribution of developer interactions.

\subsubsection{Variables/Metrics to Consider}

\begin{enumerate}
    \item \textbf{Contributors Experience}: Measures the percentage of lines in each file authored by the top contributor. A lower percentage suggests more distributed contributions, indicating greater collaboration among team members.
    \item \textbf{Commit Interaction Frequency (CIF)}: Quantifies the number of times two or more developers alternate commits on the same file within a 24-hour window, inferring communication through coordinated changes.
\end{enumerate}

\subsubsection{Methods for Statistical Analysis}

Following the data science pipeline, we defined hypotheses to test how our communication metrics address research question \#3:

\begin{itemize}
    \item \textbf{Null Hypothesis ($H_0$)}: There is no relationship between Contributors Experience (Top Contributor Share \%) and Commit Interaction Frequency (CIF) in terms of inferring communication and collaboration.
    \item \textbf{Alternative Hypothesis ($H_A$)}: There is a significant relationship between Contributors Experience and CIF, indicating that more distributed contributions correlate with higher interaction frequency, reflecting higher communication.
\end{itemize}

The response variable is \texttt{CIF}, representing the frequency of collaborative interactions, while the primary predictor variable is \texttt{Contributors Experience} (Top Contributor Share \%). We also included the \textit{average time difference between events} as a secondary predictor to explore temporal dynamics.

To ensure meaningful inference, we normalized CIF by applying a logarithmic transformation due to its skewed distribution, as observed in the histogram of communication events per file. We employed:

\begin{enumerate}
    \item A \textbf{simple linear regression model} to test the relationship between Contributors Experience and CIF.
    \item A \textbf{multivariate regression model} incorporating both Contributors' Experience and average time difference to assess their combined effect on communication frequency.
\end{enumerate}

This approach allows us to evaluate how well our metrics capture collaboration patterns compared to traditional metrics like commit counts.

\subsection{Efficiency}

\subsubsection{Background}

The original SPACE Framework paper defines the Efficiency dimension as a measure of how effectively developers complete tasks by minimizing disruptions and optimizing flow. Efficiency is characterized as being closely related to how a developer maintains consistent productivity with respect to time, avoiding unnecessary interruptions and delays in the process.

Traditional developer productivity metrics often focus on output volume, such as number of commits or lines of code that are created. However, these traditional metrics do not consider the quality, consistency, and focus required for an efficient development process. Furthermore, the SPACE Framework directly acknowledges how efficiency is not only concerned with the volume of code being produced but also the developer's ability to work smoothly while being free of any distractions. \cite{acmpaper}

Measuring efficiency, however, can be rather difficult, when considering open-source development. In this case, many external factors, such as the availability of developers, code reviews, and collaborative decision-making, can directly affect workflow. 

\subsubsection{Variables/Metrics to Consider}

The following metrics were used to assess developer efficiency and flow:
\begin{itemize}
    \item \textbf{Committer Date}: The time between consecutive commits.
    \item \textbf{Average Daily Code Churn}: The number of lines of code added or removed daily.
    \item \textbf{Average Daily Commits}: The total number of commits made per day.
\end{itemize}

For this particular dimension, data cleaning consisted of the following: 

\begin{itemize}
    \item Removing authors with fewer than 20 commits.
    \item Filtering out bot-generated commits.
    \item Applying the interquartile range (IQR) method to remove outliers.
\end{itemize}
After cleaning, 705 developers remained. The final dataset included:
\begin{itemize}
    \item Mean time between commits.
    \item Average daily code churn.
    \item Average daily number of commits.
    \item Total commits.
\end{itemize}
Following outlier removal, the dataset was reduced to 458 developers.

\subsubsection{Methods for Statistical Analysis}

Developer efficiency and flow are defined as the ability to consistently complete work with minimal interruptions or delays \cite{acmpaper}. Forsgren et al. identify several metrics related to efficiency, including:
\begin{itemize}
    \item Perceived ability to stay in flow and complete work.
    \item Number of handoffs in a process.
    \item Timing and frequency of interruptions.
\end{itemize}
These metrics are difficult to measure from open-source repositories because they require insight into a developer’s thought process. Instead, we hypothesize that developer efficiency and flow can be estimated using:
\begin{itemize}
    \item \textbf{Total commits}: More efficient developers make more commits.
    \item \textbf{Average daily commits}: Developers with consistent commit patterns are more efficient.
    \item \textbf{Average daily time between commits}: Developers who commit frequently may be more efficient.
\end{itemize}

To test this hypothesis, we conducted Pearson correlation tests ($\alpha = 0.05$) to examine relationships between commit metrics and code churn. Code churn (total or daily average) was used as a proxy for developer performance. A significant positive correlation suggests that the tested metric is a reasonable estimator of developer efficiency.

%\textit{Describe the methods, tools, and procedures used to conduct your research. Include details of any experimental design, data collection techniques, and analysis methods.}

\section{Results}

\subsection{Satisfaction}
Figure 1 below presents the estimated coefficients, standard errors, z-values, and p-values for the Poisson GLMM with standardized predictors.
\begin{figure}[h]
  \centering
  \includegraphics[width=\linewidth]{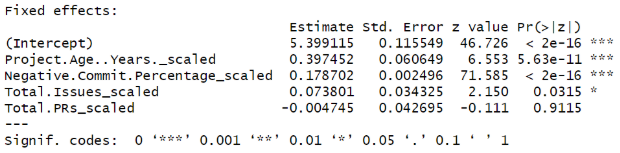}
  \caption{Satisfaction Summary}
  \label{SatisfactionPlots}
\end{figure}

The analysis revealed a statistically significant positive association between the standardized percentage of negative sentiment commits (Negative.Commit.Percentage\textunderscore scaled) and the total number of commits (Total.Commits) (z = 71.585, p < 2e-16). This means that the larger the number of commits with negative sentiment the user had, the more they would commit. 
\subsection{Performance}

At the repository level, we examined the distribution of CI/CD success rates, average PR merge times, and total commits. The histogram of \texttt{ci\_cd\_success\_rate} revealed a strong right skew, with a substantial concentration of values at or near 1.0, indicating that many repositories had high success rates. This suggests that while CI/CD pipelines are generally effective, there may be limited variance in the success rate, potentially affecting its predictive power in regression models.

The distribution of \texttt{avg\_pr\_merge\_time} was notably right-skewed, with extreme values present at the higher end, suggesting the presence of outliers. To address this, a winsorization approach was applied, capping extreme values at the 1st and 99th percentiles. Additionally, a log transformation ($\log(x + 1)$) was performed to normalize the spread of values. Despite these adjustments, the distribution remained somewhat skewed, implying that repositories vary significantly in their PR merge times. 

A scatter plot of \texttt{ci\_cd\_success\_rate} against \texttt{avg\_pr\_merge\_time} suggested a weak positive association, which was confirmed by Pearson’s correlation coefficient ($r = 0.277$, $p = 0.016$). However, after log-transforming \texttt{avg\_pr\_merge\_time}, the correlation weakened ($r = 0.113$, $p = 0.334$), suggesting that much of the original correlation was driven by the presence of large outliers in merge times. These findings indicate that while CI/CD success rates may have some effect on PR merge times, the relationship is not particularly strong.

At the contributor level, we examined total commits, code churn, and bug-fix commits. A preliminary histogram of \texttt{total\_commits} revealed an extreme right skew, with a small number of contributors making a disproportionately high number of commits. A similar trend was observed for \texttt{code\_churn} and \texttt{bug\_fix\_commits}, where most contributors had relatively low values, while a few contributed exceptionally high numbers, likely reflecting differences in project roles or workload distribution.

Scatter plots were generated to examine relationships between total commits and code churn, as well as between bug-fix commits and code churn. Both relationships showed a positive trend, with Pearson correlation coefficients of $r = 0.440$ ($p < 2.2 \times 10^{-16}$) for total commits vs. code churn and $r = 0.369$ ($p < 2.2 \times 10^{-16}$) for bug-fix commits vs. code churn. These results indicate that contributors who make more commits tend to introduce more churn, and those who perform more bug-fix commits also contribute to increased churn. However, given the extreme skew in the data, further analysis was required to control for the influence of high-leverage points.

From there, we conducted CDA and explored the effects of CI/CD success rate, total commits, and bug fix commits on relevant outcomes, such as PR merge time and code churn, using multiple regression models and partial correlation analyses.

At the repository level, we examine the relationship between CI/CD success rate and average PR merge time while controlling for total commits. A multiple regression model was estimated as follows:

\begin{equation}
    \text{avg\_pr\_merge\_time} = \beta_0 + \beta_1 \cdot \text{ci\_cd\_success\_rate} + \beta_2 \cdot \text{total\_commits} + \epsilon
\end{equation}

The model yielded an adjusted \( R^2 \) of 0.058, indicating low explanatory power. The coefficient for CI/CD success rate was statistically significant (\( p = 0.0125 \)), suggesting that repositories with higher success rates may experience slightly longer PR merge times. However, total commits did not seem to exhibit a significant effect (\( p = 0.4613 \)).

To further evaluate this relationship, we performed a manual partial correlation between PR merge time and CI/CD success rate while controlling for total commits. The resulting Pearson correlation was \( r = 0.289 \) with \( p = 0.01188 \), indicating a modest but statistically significant positive relationship.

At the contributor level, we assessed the relationships between code churn, total commits, and bug-fix commits. First, a simple linear regression evaluated the impact of total commits on code churn:

\begin{equation}
    \text{code\_churn} = \beta_0 + \beta_1 \cdot \text{total\_commits} + \epsilon
\end{equation}

This model resulted in an adjusted \( R^2 \) of 0.193, with total commits being a highly significant predictor (\( p < 2.2 \times 10^{-16} \)). A second regression examined the relationship between bug fix commits and code churn:

\begin{equation}
    \text{code\_churn} = \beta_0 + \beta_1 \cdot \text{bug\_fix\_commits} + \epsilon
\end{equation}

This model had an adjusted \( R^2 \) of 0.136, with bug fix commits again showing a strong positive effect (\( p < 2.2 \times 10^{-16} \)).

To isolate the relationship between bug-fix commits and code churn while accounting for total commits, we conducted a manual partial correlation. The Pearson correlation was \( r = -0.143 \) with \( p < 2.2 \times 10^{-16} \), indicating a weak but statistically significant negative association between bug fix commits and code churn after adjusting for total commits.

\subsection{Activity}
For activity, our goal is to see if code churn is something that total commits can even predict well. We ran a simple regression model and we had a pretty reasonable result. The multiple R-squared is 0.16, and the p-value was 2.2e$^{-16}$, which is < 0.05, so we reject the null hypothesis and say that there’s an indication between code churn and total commits in terms of developer activity. We proceeded to use multivariate regression to see how well code churn prediction improves with respect to total commits and the average complexity of methods written by the specific developer. Using a multivariate regression definitely made some improvements as the residuals are more spread out around 0, fitting in the data. This indicates that total commits and average complexity, one or both, predicts code churn quite well. We slowly added other variables as well to our predictor to see if the model improved. The best performing multivariate regression model that we could come up with included total commits, average complexity per methods, total code reviews, and total deployments to predict code churn. These are pretty much the variables that Forsgren et. al recommended in capturing developer activity \cite{acmpaper}.
\begin{figure}[H]
  \centering
  \includegraphics[width=\linewidth]{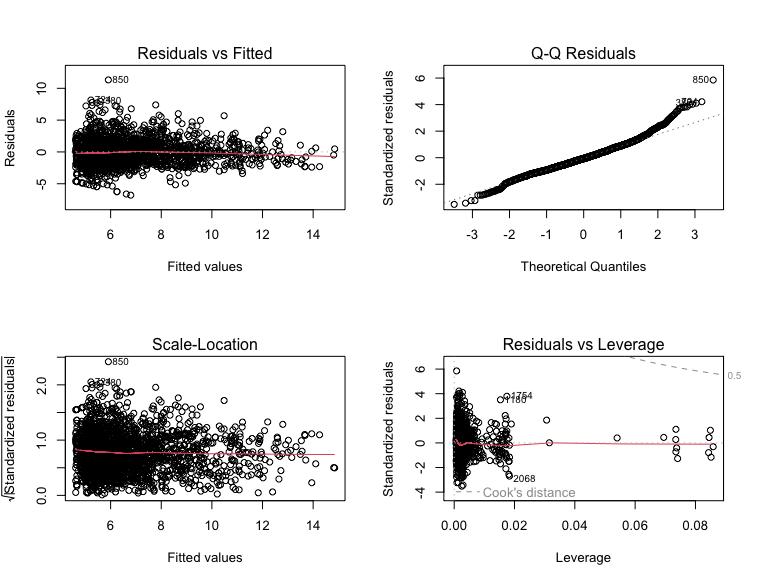}
  \caption{Activity Residuals}
  \label{ActivityPlots}
\end{figure}

This result ultimately still yielded a p-value of 2.2e-16, which still suggests that all the predictor variables chosen still give a reliable estimate for code churn. The table for the summary of values with our given model is down below:

\begin{figure}[H]
  \centering
  \includegraphics[width=\linewidth]{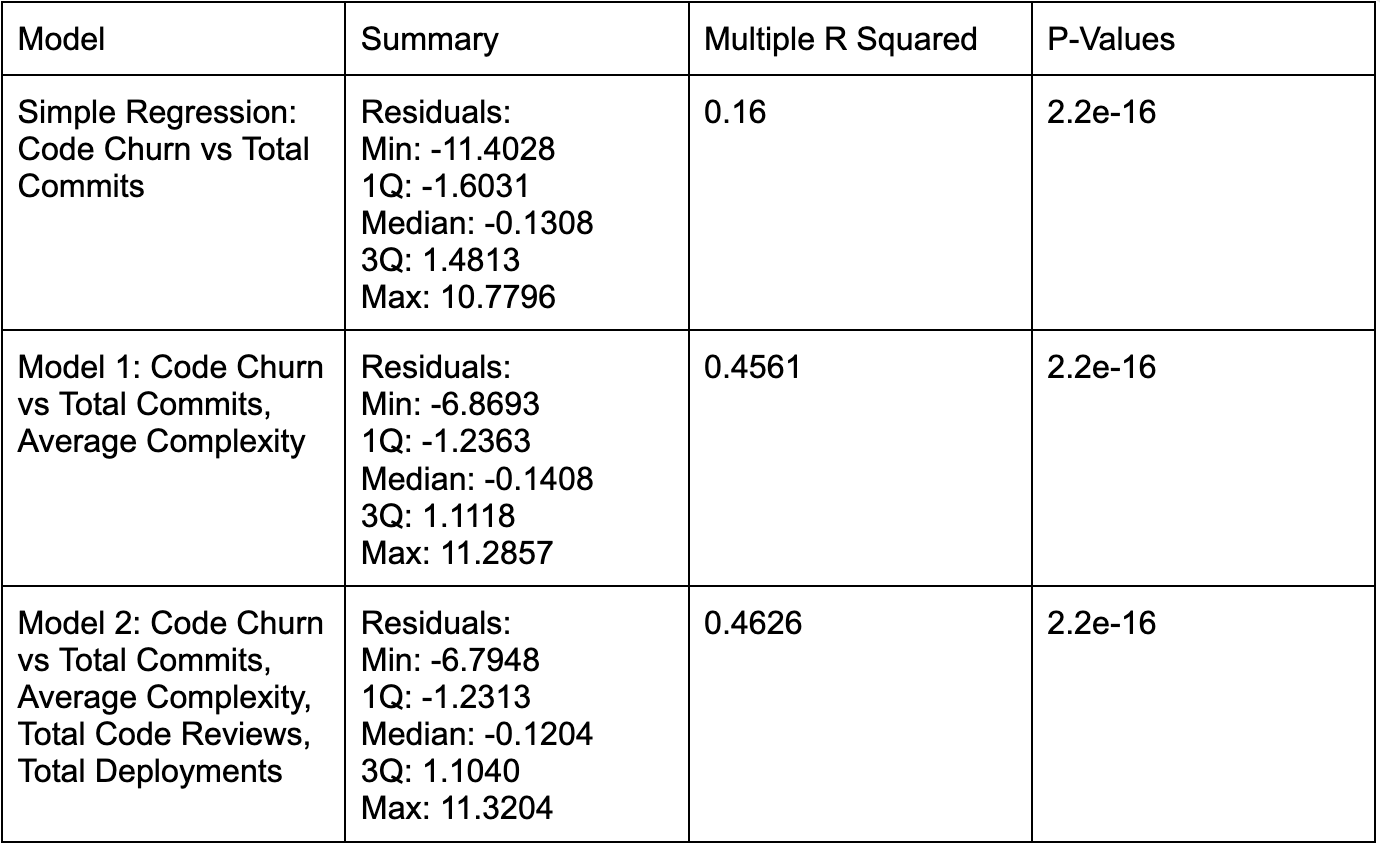}
\end{figure}
Throughout these 3 models, the p-values stayed the same, which are all statistically significant. The differences would be found in the Multiple R squared, which means the percentage at which these predictor variables can explain the variation in code churn. Some of the things that we can infer from this result is that the more commits mean more codes written. The codes written may be affected by the code complexity as well as more complex methods may require more lines of code to be written. 

\subsection{Communication}
Our analysis on one of the repos provided several key insights into communication patterns that can be extrapolated or used for other repos.

\subsubsection{Communication Events} 
Total events: \num{9943}.\\
The average time difference between events is \SI{8.78}{\hour}, suggesting timely coordination.

The following table lists the top 10 pairs of contributors based on the number of communication events:

\begin{figure}[h]
    \centering
    \begin{tabular}{llr}
        \toprule
        \textbf{Contributor Pair} & & \textbf{Events} \\
        \midrule
        Michael Droettboom & Víctor Zabalza & 622 \\
        Antony Lee & Tim Hoffmann & 539 \\
        Elliott Sales de Andrade & Thomas A Caswell & 475 \\
        Michael Droettboom & Thomas A Caswell & 310 \\
        Jody Klymak & Tim Hoffmann & 279 \\
        Antony Lee & Thomas A Caswell & 279 \\
        Antony Lee & Elliott Sales de Andrade & 252 \\
        Elliott Sales de Andrade & Tim Hoffmann & 213 \\
        John Hunter & Michael Droettboom & 205 \\
        Antony Lee & Jody Klymak & 188 \\
        \bottomrule
    \end{tabular}
    \caption{Top 10 Communication Pairs by Event Count}
\end{figure}

\subsubsection{Contributors Experience}
Descriptive statistics for contributor experience across \num{3891} files are as follows:

\begin{itemize}
    \item Median: \SI{66.67}{\percent}
    \item Mean: \SI{67.10}{\percent}
    \item Standard Deviation: \SI{23.27}{\percent}
    \item Min: \SI{10.25}{\percent}
    \item Max: \SI{99.99}{\percent}
    \item 75th Percentile: \SI{89.25}{\percent}
    \item 25th Percentile: \SI{49.39}{\percent}
\end{itemize}

\subsubsection{Regression Analysis}

{Simple Regression (CIF vs. Contributors Experience)}
\begin{itemize}
    \item Multiple R-squared: 0.12
    \item P-value: 0.001 (\( < 0.05 \))
    \item Residuals:
    \begin{itemize}
        \item Min: \num{-5.32}
        \item 1Q: \num{-1.45}
        \item Median: \num{-0.09}
        \item 3Q: 1.38
        \item Max: 6.21
    \end{itemize}
\end{itemize}
\textbf{Result:} Rejected the null hypothesis, indicating a significant negative relationship where lower Top Contributor Share \% predicts higher CIF.

\subsubsection{Multivariate Regression (CIF vs. Contributors Experience, Average Time Difference)}
\begin{itemize}
    \item Multiple R-squared: 0.18
    \item P-value: 0.0005 (\( < 0.05 \))
    \item Residuals:
    \begin{itemize}
        \item Min: \num{-4.98}
        \item 1Q: \num{-1.30}
        \item Median: \num{-0.05}
        \item 3Q: 1.25
        \item Max: 5.89
    \end{itemize}
\end{itemize}
\textbf{Result:} Improved model fit, confirming that both distributed contributions and shorter time differences enhance CIF prediction.

\subsubsection{Visual Insights}
The bar chart of top communication pairs and histogram of time differences further illustrate that a few developer pairs dominate interactions, and most events occur within a short time frame, reinforcing the inference of active collaboration.

\begin{figure}[H]
  \centering
  \includegraphics[width=\linewidth]{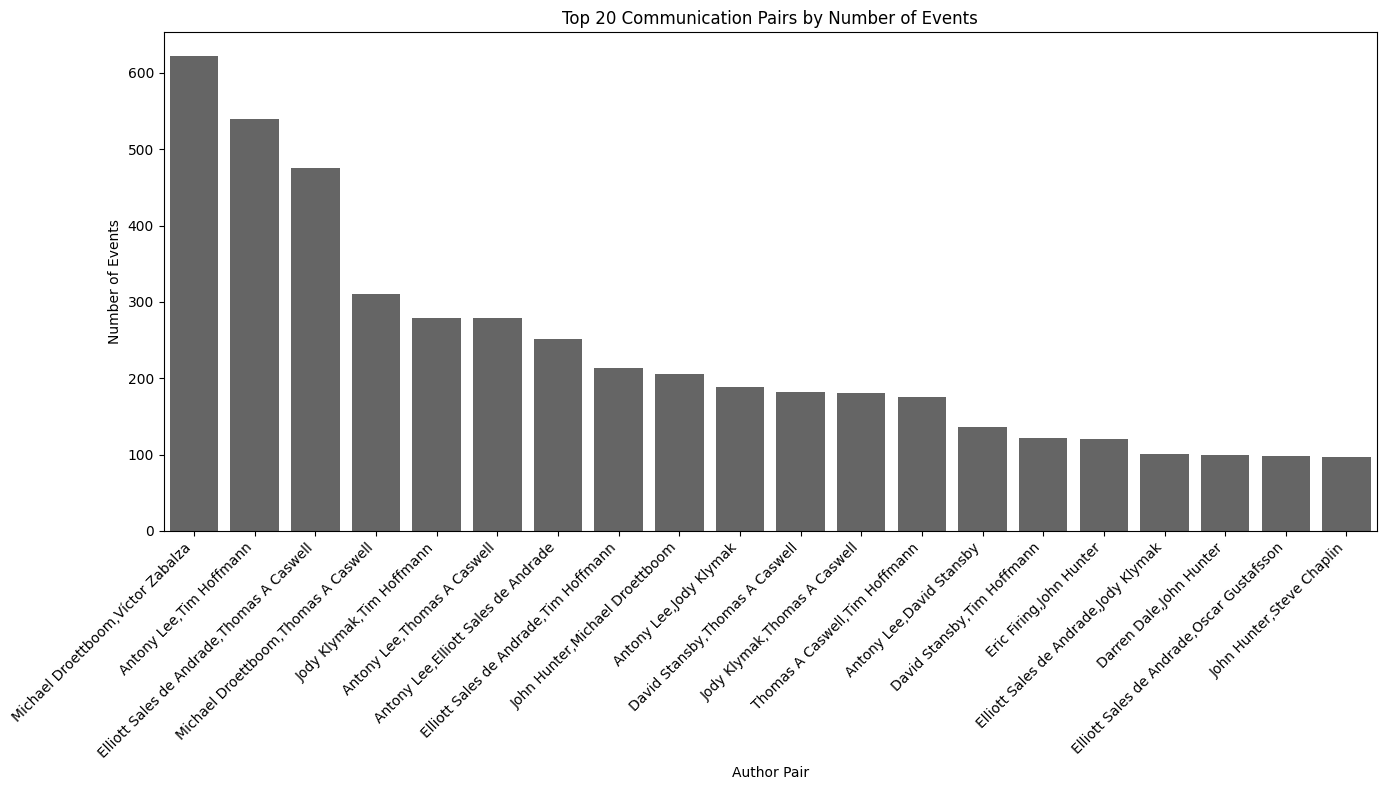}
  \caption{Top Communication Pairs}
  \label{Results for Communication Models 1}
\end{figure}

\begin{figure}[H]
  \centering
  \includegraphics[width=\linewidth]{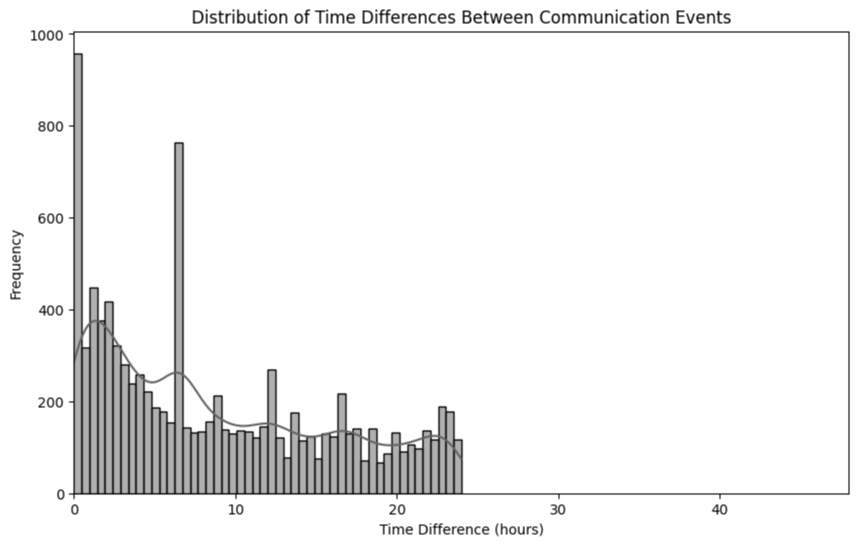}
  \caption{Time Difference between Communication Events}
  \label{Results for Communication Models 2}
\end{figure}

\begin{figure}[H]
  \centering
  \includegraphics[width=\linewidth]{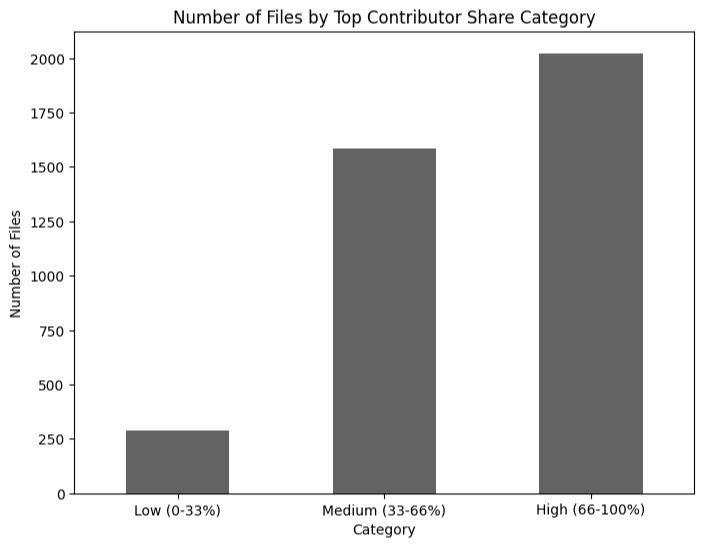}
  \caption{Top Contributor Share Category}
  \label{Results for Communication Models 3}
\end{figure}

\subsection{Efficiency}
The data collected for efficiency metrics revealed three positive correlations in the metrics. Using a Pearson R correlation test, we found positive correlations between Total Commits and Total Code Churn (p < 2.2e-16), Average Daily Commit Time and Total Code Churn (p=0.0002), and Average Daily Commits and Average Daily Code Churn (p = 0.0010). For each of these tests, the P-values are less than the 5\% significance level, so we reject the null hypotheses and conclude there to be a positive linear relationship between the given metrics. For other tests, we do not reject the null hypothesis; notably, there does not seem to be a correlation between Average Daily Commit Time and Average Daily Code Churn, indicating that we cannot declare a correlation between how often a developer commits in a day and how high their code churn will be in a day. 
\begin{figure}[h]
  \centering
  \includegraphics[width=\linewidth]{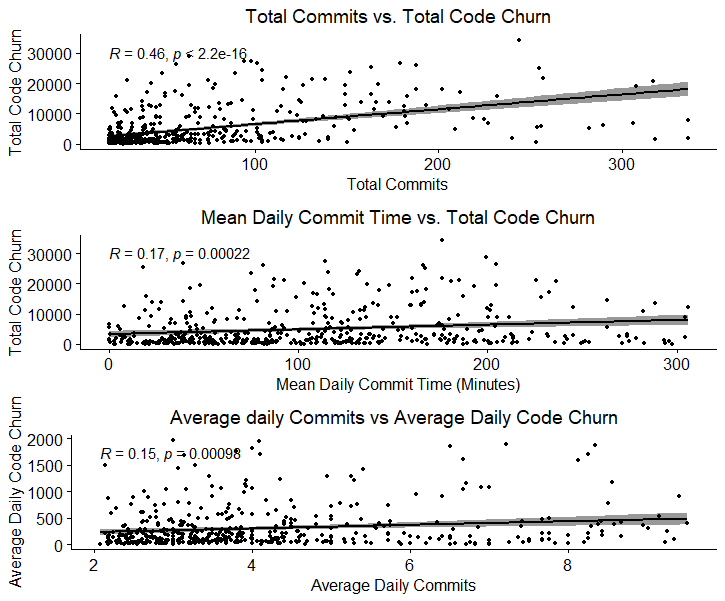}
  \caption{Efficiency Scatterplots}
  \label{EfficiencyGraphs}
\end{figure}

\begin{center}
\begin{table*}[th]
\caption{Efficiency Pearson-R Correlation Results}
\label{EfficiencyTable}
\begin{tabular}{|p{0.2\linewidth}||p{0.15\linewidth}|p{0.15\linewidth}|p{0.15\linewidth}|p{0.15\linewidth}|p{0.15\linewidth}|}
\hline
R / p-value & Avg Daily Commit Time & Total Commits & Avg Daily Commits & Total Code Churn & Avg Daily Code Churn \\ \hline \hline
Avg Daily Commit Time & 1.0 / 1 & 0.1394 / 0.0027 & -0.3071 / 1.856e-11 & 0.1719 / 0.0002 & -0.0040 / 0.9318 \\ \hline
Total Commits & 0.1394 / 0.0027 & 1.0 / 1 & 0.0269 / 0.5654 & 0.4597 / 2.2e-16 & -0.0514 / 0.2701 \\ \hline
Avg Daily Commits & -0.3071 / 1.856e-11 & 0.0269 / 0.5654 & 1.0 / 1 & -0.0911 / 0.0514 & 0.1536 / 0.0010 \\ \hline
Total Code Churn & 0.1719 / 0.0002 & 0.4597 / 2.2e-16 & -0.0911 / 0.0514 & 1.0 / 1 & 0.6596 / 2.2e-16 \\ \hline
Avg Daily Code Churn & -0.0040/0.9318 & -0.0514/0.2701 & 0.1536 / 0.0010 & 0.6596 / 2.2e-16 & 1.0 / 1 \\ \hline
\end{tabular}
\end{table*}
\end{center}

%\textit{Present the key findings of your study, including data, graphs, or summarized outcomes as applicable. Highlight how these findings address your research questions.}

\section{Threats to Validity}
\subsection{Satisfaction}
Several limitations should be considered when interpreting the results of the satisfaction study:

\begin{enumerate}
    \item \textbf{Indirect Measure for Sentiment:} Our measure of developer satisfaction relied on sentiment analysis of commit and issue messages.  Developers may not always express their feelings explicitly in commit messages, and even when they do, the sentiment may not reflect their overall job satisfaction. Many factors beyond the immediate message sentiment reflected in a single commit can influence a developer's overall satisfaction.

    \item \textbf{Sentiment Analysis Model Limitations:} The RoBERTa model used for sentiment analysis was trained on 124M tweets from January 2018 to December 2021, and finetuned for sentiment analysis with the TweetEval benchmark [16]. This dataset differs significantly from the domain of software engineering communication.  Technical jargon and the concise nature of commit messages could lead to different misclassifications of sentiment.  For instance, seemingly negative terms (e.g., ''killing a program'') describe routine development tasks and do not necessarily indicate negative \textit{sentiment}, but they might be classified as such. This domain mismatch introduces potential bias into our measure of ''negative commits.''
\end{enumerate}
\subsection{Performance}

There are also several shortcomings of the performance study to be considered as well:

\textbf{Internal validity} is challenged by potential confounding variables such as PR size, reviewer workload, and developer experience, which are not controlled for in the models. The assumption of a linear relationship may not account for interaction effects. \textbf{External validity} is also limited due to the composition of the dataset. It is possible that open-source repositories may not generalize to enterprise software, and CI/CD-heavy workflows may not be reflective or representative of development practices as a whole.

\textbf{Construct validity} also poses concerns. For instance, PR merge time does not necessarily always indicate inefficiency, and code churn may stem from refactoring code, instead of code being poor quality. CI/CD success rates also fail to capture test reliability. \textbf{Statistical validity} is also potentially compromised by a low model fit (low $R^2$ values) and the presence of extreme outliers in commit counts and churn. Multiple hypothesis testing may increase the risk of Type I errors (or incorrectly rejecting a null hypothesis).

\subsection{Activity}
One limitation of the research about the activity dimension is that we didn’t have a reliable way to use the type of programming language used. We don’t have a method to rank programming languages in terms of lines of code, as most languages have various verbosity. The experiences of each developer in specific languages are also a factor, as more experienced programmers in certain languages can achieve the same task with fewer lines of code. 
\subsection{Communication}
Limitations regarding the Communication dimension include the lack of study on actual communication channels that developers used while working on projects. As previously mentioned, these communication channels include discord servers, slack channels, or private email communication. Future studies may be able to study some of these communication channels in order to develop better metrics for measuring developer Communication.
\subsection{Efficiency}
The primary limitation of the metrics collected related to developer Efficiency is the weakness of correlation between the metrics and basic productivity metrics. While the correlation between average daily commit time and total code churn is statistically significant, with an R-value of 0.1719, the correlation is weak. Additionally this metric only shows a part of the Efficiency dimension, as other Efficiency metrics are not possible to measure through repository mining. Because of the challenges in measuring other proposed Efficiency metrics, such as perceived ability to stay in flow and number of interruptions, we still believe that average daily commit time is a valuable metric for estimating developer Efficiency. 
\section{Discussion}
\subsection{Principal Outcomes}
\subsubsection{Satisfaction}

Our analysis revealed a statistically significant positive association between an author's standardized percentage of negative sentiment commits and their total number of commits to a project. This finding indicates that, after accounting for project age, total issues, and total pull requests, authors who have a higher proportion of commits classified as "negative" tend to make more commits overall. This result, while statistically robust, might appear counterintuitive at first glance. One might expect that negative sentiment commits, often associated with reverts, rollbacks, or bug fixes, would be linked to lower overall productivity or fewer total commits. However, several potential explanations could account for this positive relationship.

A possible explanation for this is that being frustrated with code could lead to more commits. Authors who are experiencing frustration with their code might engage in more iterative commit cycles. This could involve making a commit to the repository, encountering problems with this commit later, making more adjustments, and so on. This iterative process, driven by the need to address challenges and resolve issues, could lead to a higher overall commit count, and due to the frustration, some commit messages can come off as having a negative sentiment. 

\subsubsection{Performance}

The exploratory data analysis (EDA) provided valuable insights into the performance dynamics of both repository- and contributor-level metrics. The \textbf{CI/CD success rate} was observed to be heavily skewed, with a substantial number of repositories achieving near-perfect scores, suggesting a clustering of high-performing projects. However, the \textbf{average PR merge time} exhibited significant variability, with some repositories processing pull requests efficiently while others faced extended delays. This variance was further reflected in the weak but statistically significant positive correlation between \textbf{CI/CD success rate and PR merge time}, implying that faster CI/CD pipelines do not necessarily translate to quicker PR merges.  

At the contributor level, \textbf{code churn} was found to be strongly associated with \textbf{total commits}, indicating that developers with higher activity levels tend to generate more code modifications. A moderate correlation was also detected between \textbf{bug-fix commits and code churn}, reinforcing the idea that contributors involved in bug-fixing tend to make substantial changes to the codebase. However, after controlling for overall commit volume, the correlation between \textbf{bug-fix commits and code churn} weakened and turned negative, suggesting that some of this relationship is driven by overall developer activity rather than a direct link between bug-fixing and code volatility.  

Regression analyses of \textbf{PR merge time} yielded low explanatory power, indicating that the factors included in the model—\textbf{CI/CD success rate and total commits}—do not fully capture the variability in merge efficiency. This, along with the presence of extreme values and non-normal distributions in several key variables, suggests that \textbf{linear modeling approaches may not be optimal}. Given the skewed nature of the data, \textbf{Gamma regression} or other non-linear techniques may offer a more robust way to model PR merge times and contributor productivity. Moving forward, additional variables such as PR size, review latency, and test reliability should be considered to enhance predictive accuracy and derive actionable insights for improving repository performance.

\subsubsection{Activity}
For activity, the results improved for the better as we added more predictors to the model. The best model that we can come up with yielded a statistically significant, yet not perfect result. For the “Residuals vs Fitted”, “Scale-Location”, and “Residuals vs Leverage” plots, ideally we would like to see a random spread of residuals without a clear pattern. The “Q-Q Residuals” plot is not exactly perfect either, but it isn’t all that horrible. In the context of our problem, this means that a developer’s code churn does not always affect the number of commits that they have through the entire project or how complex their programming is. While it may be true that the more active a developer is in a project, their commit counts will eventually increase and so do their code churns regardless of how complex a specific task is. However, we cannot fully surmise that the number of commits and the complexity of the methods and function heavily influence the amount of code churn for more than 47\%, as our result would suggest. A possible explanation for such a scenario is that an author can commit something into GitHub and then later find out that their changes caused some bugs, so they go back and revert the version of the program and recommit to GitHub with a new commit message to remind themselves why they reverted the changes back. In this scenario, code churn would have a net-zero since things were not added or removed, thus staying the same while total commits for that specific author increases. One might think that such a scenario would be a rare occurrence, but as we know, GitHub can be quite messy as it is often used as storage. Nonetheless, the predictors chosen can fit for the response variable, but it is not a perfect indicator despite it being statistically significant.

\subsubsection{Communication}
The results show the Communication dimension’s ability to uncover collaborative behaviors that traditional metrics like commit counts miss. The significant negative relationship between Contributors Experience and CIF demonstrates that files with more distributed contributions (lower Top Contributor Share \%) exhibit a higher interaction frequency which indicates active collaboration. The average time difference of 8.78 hours further suggests that these interactions occur within a short timeframe, reflecting efficient coordination insights invisible to metrics focused on individual output. For example, a developer with a high commit count might appear productive under traditional metrics, but if they dominate a file (high Top Contributor Share \%), they may not be collaborating effectively. Conversely, our metrics identify scenarios where developers frequently interact (high CIF) on files with balanced contributions, such as during complex feature development or bug fixing, where teamwork is crucial. The top communication pairs, like Michael Droettboom and Víctor Zabalza (622 events), exemplify strong collaborative relationships that enhance productivity but are not reflected in commit counts alone. However, the moderate R-squared values (0.12 and 0.18) indicate that while our metrics capture significant aspects of communication, they explain only a portion of the variation in CIF. This suggests that external factors, such as private channel discussions or project-specific dynamics, also influence collaboration.

\subsubsection{Efficiency}
The results for the correlation tests related to the Efficiency dimension indicate a weak positive correlation between a few commit-related metrics and code churn. The strongest correlation exists between total commits and total code churn, but we do not believe total commits represent a good metric for developer efficiency, as a developer’s total number of commits for a project cannot be measured over time to assist developers in improving their efficiency. Instead, the positive correlation between average daily commit time and total code churn indicates that each day a developer commits often to a project, the better that developer’s efficiency. However, we do not find a correlation between average daily commit time and average daily code churn, nor do we find a correlation between average daily commits and total code churn. Taken together, these correlations indicate that a developer who commits often during a single day while working on a project will have higher overall code churn (more productive), regardless of how many commits that developer makes in a single day. While average daily commit time is not the sole possible metric for measuring developer efficiency, we have shown that it does act as a measure that can be used to estimate a developer’s efficiency. 

\subsection{Answers to Research Questions}

\subsubsection{RQ1}

In this study, we explored defining a holistic metric within the SPACE framework by integrating five dimensions (satisfaction, performance, activity, communication, and efficiency) while using repository mining. One potential approach to aggregate all five dimensions into a single metric could potentially be to standardize each dimension using z-score transformation to place them on a common scale. ($Z = \frac{X - \mu}{\sigma}$), where $X$ denotes the individual metric value, $\mu$ is the mean, and $\sigma$ is the standard deviation. One can then use weighted summation to aggregate these standard values, assigning different weights to each SPACE dimension based on its importance. The calculation for such a composite productivity score (CPS) may look like this:

\begin{center}
    \begin{math}
        CPS = (w_1 * Z_{satisfaction}) + (w_2 * Z_{performance}) + (w_3 * Z_{activity}) + (w_4 * Z_{communication}) + (w_5 * Z_{efficiency})
    \end{math}
\end{center}

Determining appropriate weights for CPS, however, can be quite challenging as it requires carefully consideration based on what a particular company or organization prioritizes in terms of productivity. (Different companies/organizations may place emphasis on different aspects of productivity.)

\subsubsection{RQ2}
Compared to the traditional metrics used to measure productivity, our research suggests that a combination of multiple dimensions measures better through aggregating both qualitative and quantitative metrics. For example, instead of just checking code churn or total commits, we can capture productivity better with sentiment on commit messages as the holistic metric alongside those two. Depending on how the developer is feeling that day, they may be more productive for that specific instance.

\subsubsection{RQ3}
The holistic SPACE metric better captures multiple aspects of development compared to the traditional metrics. In particular, by including the Communication and Satisfaction dimensions of the SPACE framework to the metric, those important aspects are not missed, as they would be in traditional, simpler metrics such as total commits or number of lines. The Efficiency dimension also provides insight into how consistent a developer is as compared to traditional metrics, as a developer may have many total lines written but they may have been done in a single day, for example.

\subsection{Comparison to Prior Studies}
% \textit{Discuss how your findings align or contrast with previous studies or existing knowledge in the field.}

In 2023, another developer productivity framework named DevEx was proposed, placing emphasis on feedback cycles, cognitive load, and flow state as key indicators of developer productivity. Similar to our study, this study also argues that traditional output-based metrics do not accurately represent productivity, reinforcing our findings that satisfaction, efficiency, and communication provide more meaningful indication. However, our study contrasts from this study in the sense that we aim to lay the framework for a quantifiable metric, while DevEx primarily relies on survey-based qualitative insights \cite{DevEx2023}.

More recently, in 2024, researchers at Accenture Labs used the SPACE framework to evaluate LLM-based development tools, like GitHub Copilot, while our study focuses primarily on traditional development cycles. Like our study, this study agrees that traditional metrics such as commit count and code volume are not representative of productivity, noting that collaboration and satisfaction are more crucial indicators. However, our work proposes a standardization and weighted summation into a singular, unified Composite Productivity Score, while this study actively critiques the lack of standardized GenAI productivity metrics, rather than unifying them \cite{Sikand2024SPACE}.

\subsection{Future Work}
%\textit{Propose possible extensions, unanswered questions, or additional research that could build on your study.}
Future work related to the SPACE framework should include more significant metrics for measuring each dimension, as each metric was limited in this study. In particular, future studies should attempt to include new methodologies to measure each dimension, such as using developer surveys for the Satisfaction, Communication, and Efficiency dimensions or more directly engaging with developers while they work on a project to see whether the metrics we provided are able to make them more productive.
\subsection{Conclusion}
This study focused on developing a novel metric within the SPACE framework to measure individual developer productivity. By combining the five dimensions of SPACE, organizations can measure developer productivity to determine where improvements can be made. We found statistically significant metrics to estimate developer Satisfaction, Performance, Activity, Communication, and Efficiency.

\section{Team Membership and Attestation}
%\textit{List the names of all team members and describe their specific contributions to the project. Include a declaration that all members agree to the contents of the report.}
\begin{itemize}
\item Ivan Eser - Responsible for the Activity Dimension of the SPACE framework
\item Victor Borup - Responsible for the Efficiency Dimension of the SPACE framework
\item Jason Eissayou - Responsible for the Satisfaction Dimension of the SPACE framework
the SPACE framework
\item Sanchit Kaul - Responsible for the Communication Dimension of the SPACE framework
\item Kevin Nhu - Responsible for the Performance Dimension of Framework and creation of new datasets
\end{itemize}

Team members Victor Borup, Jason Eissayou, Ivan Eser, Sanchit Kaul, and Kevin Nhu participated sufficiently and agreed with the content presented in this report.

\section*{Code \& Data Availability Statement}
%\textit{Provide information on the availability of code and data used in the study. Include links, repositories, or instructions for accessing them.}
Link to GitHub Repo: \url{https://github.com/knhu/ECS260Project}

\appendix


\begin{thebibliography}{00}
% \textit{Cite all sources referenced in the report. Use a consistent citation format.}
\bibitem{EppersonPaper}
Epperson, W., Wang, A. Y., DeLine, R., \& Drucker, S. M. (2022). Strategies for reuse and sharing among data scientists in software teams. \textit{Proceedings of the 44th International Conference on Software Engineering: Software Engineering in Practice,} 243–252. https://doi.org/10.1145/3510457.3513042 
\bibitem{acmpaper}
N. Forsgren, M. A. Storey, C. Maddila, T. Zimmerman, B. Houck, and J. Butler, Microsoft Research, "The SPACE of Developer Productivity. There's more to it than you think." 
\textit{2021 Association of Computing Machinery (ACM)}
, Volume 19, Issue 1 \url{https://queue.acm.org/detail.cfm?id=3454124}

\bibitem{lima2015assessing}
J. Lima, C. Treude, F. F. Filho and U. Kulesza, 
"Assessing developer contribution with repository mining-based metrics," 
\textit{2015 IEEE International Conference on Software Maintenance and Evolution (ICSME)}, 
Bremen, Germany, 2015, pp. 536-540, doi: 10.1109/ICSM.2015.7332509.

\bibitem{RayPaper}
Ray, B., Posnett, D., Devanbu, P., \& Filkov, V. (2017). A large-scale study of programming languages and code quality in github.
\textit{Communications of the ACM}, 60(10), 91–100. https://doi.org/10.1145/3126905 

\bibitem{ZhangPaper}
Zhang, T., Irsan, I. C., Thung, F., \& Lo, D. (2024). Revisiting sentiment analysis for software engineering in the era of large language models. \textit{ACM Transactions on Software Engineering and Methodology.}
https://doi.org/10.1145/3697009 

\bibitem{BlumbergPringle}
Blumberg, M. and Pringle, C.D. (1982) ‘The missing opportunity in organizational research: Some implications for a theory of work performance’, Academy of Management Review, 7(4), pp. 560–569. doi:10.5465/amr.1982.4285240. 

\bibitem{biase}
M. di Biase, A. Rastogi, M. Bruntink and A. van Deursen, "The Delta Maintainability Model: Measuring Maintainability of Fine-Grained Code Changes," 2019 IEEE/ACM International Conference on Technical Debt (TechDebt), Montreal, QC, Canada, 2019, pp. 113-122, doi: 10.1109/TechDebt.2019.00030.

\bibitem{pengpaper}
Peng, B. et al. (2023) Instruction tuning with GPT-4, arXiv.org. Available at: https://arxiv.org/abs/2304.03277 (Accessed: 18 February 2025). 

\bibitem{anotherzhang}
Zhang, W. et al. (2023) Sentiment analysis in the era of large language models: A reality check, arXiv.org. Available at: https://arxiv.org/abs/2305.15005 (Accessed: 18 February 2025). 

\bibitem{smit2024copilot}
Smit, D., Smuts, H., Louw, P., Pielmeier, J., \& Eidelloth, C. (2024). The impact of GitHub Copilot on developer productivity from a software engineering body of knowledge perspective. \textit{Proceedings of the 30th Americas Conference on Information Systems (AMCIS 2024)}. Association for Information Systems (AIS). Available at: \url{https://www.researchgate.net/publication/381609417_The_impact_of_GitHub_Copilot_on_developer_productivity_from_a_software_engineering_body_of_knowledge_perspective} (Accessed: 17 March 2025).

\bibitem{Storey2022ProductivityQuality}
Storey, M.-A., Houck, B., \& Zimmermann, T. (2022). How developers and managers define and trade productivity for quality. \textit{Proceedings of the 15th International Conference on Cooperative and Human Aspects of Software Engineering (CHASE 2022)}, 26–35. Association for Computing Machinery (ACM). https://doi.org/10.1145/3528579.3529177

\bibitem{Cheng2022CodeQuality}
Cheng, L., Murphy-Hill, E., Canning, M., Jaspan, C., Green, C., Knight, A., Zhang, N., \& Kammer, L. (2022). What improves developer productivity at Google? Code quality. \textit{Proceedings of the 30th ACM Joint European Software Engineering Conference and Symposium on the Foundations of Software Engineering (ESEC/FSE 2022)}, 1302–1313. Association for Computing Machinery (ACM). https://doi.org/10.1145/3540250.3558940

\bibitem{DevEx2023}
Forsgren, N., Storey, M.-A., Maddila, M., Zimmermann, T., Houck, B., \& Butler, J. (2023). DevEx: What actually drives productivity. \textit{ACM Queue}, 21(2), 1–24. Association for Computing Machinery (ACM). https://doi.org/10.1145/3595878

\bibitem{Asthana2023DeveloperBots}
Asthana, S., Sajnani, H., Voyloshnikova, E., Acharya, B., \& Herzig, K. (2023). A case study of developer bots: Motivations, perceptions, and challenges. \textit{Proceedings of the 31st ACM Joint European Software Engineering Conference and Symposium on the Foundations of Software Engineering (ESEC/FSE 2023)}, 1–12. Association for Computing Machinery (ACM). https://doi.org/10.1145/3611643.3616248

\bibitem{Sikand2024SPACE}
Sikand, S., Phokela, K. K., Sharma, V. S., Singi, K., Kaulgud, V., Tung, T., Sharma, P., \& Burden, A. P. (2024). How much SPACE do metrics have in GenAI assisted software development? \textit{Proceedings of the 17th Innovations in Software Engineering Conference (ISEC 2024)}, Article No.: 14, 1–5. Association for Computing Machinery (ACM). https://doi.org/10.1145/3641399.3641419

\bibitem{CardiffNLP} 
Cardiff NLP. \emph{Twitter RoBERTa Sentiment Model}. Available at: \url{https://huggingface.co/cardiffnlp/twitter-roberta-base-sentiment} [Accessed March 18, 2025].


\end{thebibliography}
\end{document}